\documentstyle[11pt,psfig,paspconf]{article}

\begin{document}

\title{The Submillimeter Search for Very High Redshift Galaxies}

\author{A.J. Barger, L.L. Cowie}
\affil{Institute for Astronomy, University of Hawaii,
    2680 Woodlawn Drive, Honolulu, HI 96822}
\author{E.A. Richards}
\affil{National Radio Astronomy Observatory \& University of Virginia,
    520 Edgemont Road, Charlottesville, VA 22903}

% The abstract is entered in a LaTeX "environment", designated with paired
% \begin{abstract} -- \end{abstract} commands.  Other environments are
% identified by the name in the curly braces.

% Poster authors ONLY may omit the abstract in order to gain a little
% more page space for the text of the poster.

\begin{abstract}
Identifying the optical/near-infrared (NIR) counterparts to the distant
submillimeter (submm) source population has proved
difficult due to poor submm spatial resolution.
However, the proportionality of both centimeter and 
submm data to the star formation rate suggests that 
high resolution radio continuum maps with
subarcsecond positional accuracy could be exploited to locate
submm sources. We targeted with SCUBA a sample
of micro-Jansky ($\mu$Jy) radio sources in the flanking fields of the Hubble
Deep Field selected from the uniform (8\ $\mu$Jy at 1$\sigma$)
1.4\ GHz VLA image of Richards (1999).
We find that the majority of bright ($>6$\ mJy) submm sources
have detectable radio counterparts. With the precise positions from the
radio, we also find that these submm sources are extremely faint 
in the optical and NIR ($I\gg24$ and $K=21-22$) 
and are therefore inaccessible 
to optical spectroscopy. Redshift estimates can, however, be 
made from the shape of the spectral energy distribution in the 
radio and submm. This procedure, which we refer 
to as {\em millimetric} redshift estimation,
places the bright submm population at $z=1-3$, where it forms the 
high redshift tail of the faint radio population. 
\end{abstract}

% Keywords should be included, but they are not printed in the hardcopy.


\section{Introduction}

The cumulative rest-frame far-infared (FIR) emission from all objects 
lying beyond our Galaxy, known as the cosmic FIR and 
submillimeter (submm) background, was recently
detected by the {\it FIRAS} and {\it DIRBE} experiments on the
{\it COBE} satellite (e.g.\ Puget et al.\ 1996; Fixsen et al.\ 1998). 
The discovery that this background was {\em comparable} 
to the total unobscured emission at optical/ultraviolet wavelengths immediately 
made it clear that a full accounting of the star formation history of the 
Universe could only be obtained through the resolution and detailed 
study of the individual components that make up the FIR/submm background.
%observations of the energy released through star formation and AGN radiation 
%and dust absorbed and reradiated into the rest-frame FIR

The resolution of the background at 850\ $\mu$m
became possible with the installation of the Submillimeter Common 
User Bolometer Array (SCUBA; Holland et al.\ 1999) on the
15m James Clerk Maxwell Telescope (JCMT) on Mauna Kea, and results
are now available from both blank field and lensed cluster field surveys
(Smail, Ivison, \& Blain 1997; Hughes et al.\ 1998;
Barger et al.\ 1998; Barger, Cowie, \& Sanders 1999;
Blain et al.\ 1999; Eales et al.\ 1999; Lilly et al.\ 1999).
Barger, Cowie, \& Sanders (1999) used optimal fitting techniques combined
with Monte Carlo simulations of the completeness of the source count
determinations to show that a differential source count parameterization
\begin{equation}
n(S)=N_0/(a+S^{3.2})
\end{equation}
reasonably fits both the cumulative 850\ $\mu$m source 
counts and the 850\ $\mu$m extragalactic background light (EBL) measurements.
Here $S$ is the flux in mJy, $N_0=3.0\times 10^4$ per square degree per mJy, 
and the range $a=0.4-1.0$ matches the Fixsen et al.\ (1998) and
Puget et al.\ (1996) EBL values. The 95 percent
confidence range for the index is $2.6-3.9$.
The extrapolation  suggests that the typical submm source is
about 1 mJy. The direct counts show that $\sim30$ per cent of the
850\ $\mu$m background comes from sources above 2\ mJy.

\section{Towards a redshift distribution}

The redshift distribution of the submm population is needed
to trace the extent and evolution of obscured star formation
in the distant Universe. However,
identifying the optical/NIR counterparts to
the submm sources is difficult due to the uncertainty in
the SCUBA positions. Barger et al.\ (1999b)
presented a spectroscopic survey of possible optical counterparts
to a flux-limited sample of galaxies selected from the 850\ $\mu$m
survey of massive lensing clusters by Smail et al.\ (1998).
Identifications were attempted for
all objects in the SCUBA error-boxes that were bright enough for
reliable spectroscopy; redshifts or limits were obtained for
24 possible counterparts to 14 of the 17 SCUBA sources in the
sample. The remaining three submm sources consisted of two blank fields
and a third source that had not yet been imaged in the optical when 
the spectroscopic data were obtained.
The redshift survey produced reliable identifications for
six of the submm sources:
two sets of galaxy pairs (a $z=2.8$ AGN/starburst
(Ivison et al.\ 1998) and a $z=2.6$ starburst),
two galaxies showing AGN signatures ($z=1.16$ and $z=1.06$),
and two cD galaxies (cluster contamination).
The galaxy pairs were later confirmed as the true counterparts
through the detection at their redshifts of CO emission in the 
millimeter (Frayer et al.\ 1998, 1999). 
The relative paucity of AGNs in the general field population
($<1$\%) suggests that if an AGN is identified, then the submm
emission is most likely associated with that source.
A lower limit of 20 per cent of the
submm sources in the sample show signs of AGN activity.

For the eight remaining submm sources 
for which spectroscopic data of the most probable
counterparts were obtained ($z\sim 0.18-2.11$), 
the identifications are uncertain.
As mentioned above, two of the submm detections 
in the sample have no visible counterparts in very deep 
imaging ($I\sim 26$),
and it is possible that the eight submm sources
are similarly optically faint. 
We show in the next section that this is in fact highly likely.
Such sources could either 
be at very high redshift or be so highly obscured that they are emitting 
their energy almost entirely in the submm. 
Using very deep NIR imaging of the cluster fields,
Smail et al.\ (1999) recently detected possible extremely 
red object (ERO) counterparts to two of the submm sources tentatively
identified with bright $z\sim 0.5$ spiral galaxies in Barger et al.\ (1999b).

\section{Pinpointing submm sources via radio continuum observations}

A promising alternative method for locating submm sources is through 
the use of radio continuum observations
which can be made with subarcsecond positional accuracy and resolution.
A relatively tight empirical correlation between radio continuum emission
and thermal dust emission is known to exist in nearby star forming
galaxies (Condon 1992). This FIR-radio correlation
is due to both radiation processes being connected to massive
star formation activity in a galaxy. We therefore decided to test whether 
the submm source population could be efficiently located
by targeting known radio sources with SCUBA. We chose to use
the complete Hubble flanking fields (HFF) 1.4\ GHz VLA sample 
(70 sources at $5\sigma$) of 
Richards (1999). This sample is ideally suited for our purpose due to its 
uniform (8\ $\mu$Jy at 1$\sigma$) sensitivity over the whole flanking
field region and to the availability of corresponding 
deep ground-based optical and NIR imaging (Barger et al.\ 1999a). 
Richards et al.\ (1999) found that at the $\mu$Jy level
60\% of the radio sources could be identified 
with bright disk galaxies ($I<22$) and 20\% with low luminosity
AGN. However, the remaining 20\% could not be
identified to optical magnitude limits of $I=25$.

We first made use of the precise radio positions to observe a 
complete subsample of the radio-selected objects with LRIS on the Keck~II 
10m telescope. In particular we wished to determine whether
we could identify any spectral 
features in the optically-faint radio sources. 
We were able to spectroscopically identify nearly all the 
objects in our subsample to $K<20$ (max. redshift $z\sim 1.2$), but 
we could not obtain redshifts for the fainter objects.
The latter objects might be highly obscured star forming galaxies and 
hence would be good SCUBA targets.

\begin{figure}[tbh]
\centerline{\psfig{figure=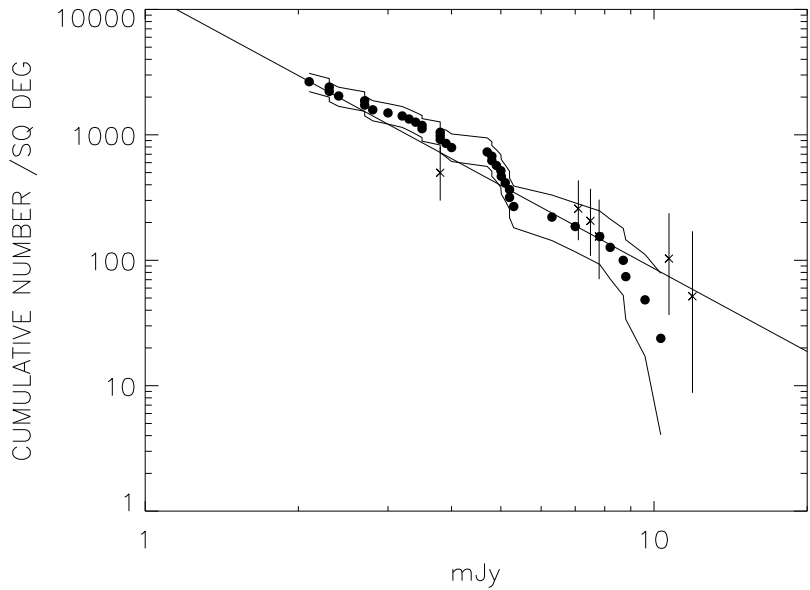,angle=90,width=4.5in}}
\caption{Comparison of radio-selected 850\ $\mu$m source
counts (X symbols with $1\sigma$ uncertainties)
from Barger, Cowie, \& Richards (in preparation; the source
below 6\ mJy is from the deep HDF-proper submm survey of
Hughes et al.\ 1998)
to the combined 850\ $\mu$m source counts (solid circles) from
blank field submm surveys (Barger, Cowie, \& Sanders 1999;
Eales et al.\ 1999; Hughes et al.\ 1998). The jagged solid
lines are $1\sigma$ uncertainties.
The solid line is a power law parameterization, $N(S)\propto S^{-2.2}$.
Radio selection is clearly an efficient means for detecting the submm source
population at bright fluxes.
} \label{fig-1}
\end{figure}

We targeted 14 of the 16 $K>21$ radio sources in the HFF sample using 
SCUBA in jiggle map mode, which enabled us to simultaneously 
observe a large fraction of the optical/NIR-bright radio sources (35/54). 
Even with relatively shallow SCUBA
observations (a 3$\sigma$ detection limit of 6\ mJy at $850\ \mu$m), 
we were able to make 5 submm detections ($>3\sigma$)
of the 14 targeted $K>21$ sources;
in contrast, none of the 35 optical/NIR-bright sources were detected.
Additionally, we detected two $>6$\ mJy sources in our observed fields
(which cover slightly more than half of the HFF) that did not have
radio detections.

Our high success rate with our targeted observations
indicates that selection in the radio 
is an efficient means of detecting the majority of the bright submm source
population. We illustrate this in Figure~1 where 
we compare our radio-selected submm source counts with
the combined source counts from published blank field submm surveys. 
An important corollary to the optical/NIR-faint radio sources being
detectable as bright submm sources is that a
large fraction of the sources in submm surveys will have
extremely faint optical/NIR counterparts and hence cannot
be followed up with optical spectroscopy.

\section{Millimetric Redshift Estimation} 

Although we are unable to obtain spectroscopic redshifts for the 
optical/NIR-faint radio-selected submm sources, we can use the shape
of the spectral energy distribution (SED) in the radio and submm 
to obtain millimetric redshift estimates.
The slope of the
SED changes abruptly at frequencies below $10^{11}$\ Hz when
the dominant contribution to the SED changes from thermal dust emission
to synchrotron radio emission. Carilli \& Yun (1999) suggested that the
presence of this spectral break would enable the use of the 
submm-to-radio ratio as a redshift estimator. 
Figure~2 shows how visual redshift 
estimates for the radio-selected submm sources can be made from a plot 
of radio flux, $S_{1.4\ {\rm GHz}}$, versus submm flux, $S_{353\ {\rm GHz}}$, 
for a range in luminosities of the prototypical 
ultra-luminous infrared galaxy (ULIG) Arp~220.
All of the submm sources detected in this survey fall in the redshift
range $z=1-3$, consistent with the redshifts
of the lensed submm sources (Barger et al.\ 1999b).

\begin{figure}[tbh]
\centerline{\psfig{figure=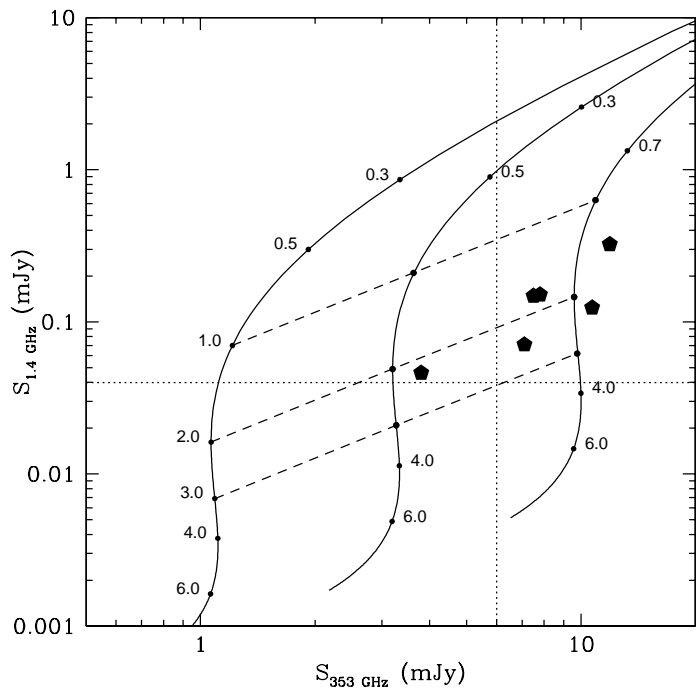,width=3.8in}}
\caption{Radio flux versus submm flux for a redshifted Arp~220 scaled
with luminosity
by factors of $1/3$, 1, and 3. Redshift labels are marked on the curves.
Dashed lines connect the curves at $z=1$, 2, and 3.
The dotted horizontal line marks the 5$\sigma$ flux limit of 0.04\ mJy
for the 1.4\ GHz radio sample. The dotted vertical line marks a
$3\sigma$ detection limit of 6\ mJy in the submm. The five 
solid pentagons at $S_{353}>6$\ mJy
are distant ULIGs detected in our radio-selected submm survey; 
the faint submm source is from Hughes et al.\ (1998).
} \label{fig-2}
\end{figure}

For $z<3$, we can express the Arp~220 predicted flux ratio as
\begin{equation}
S_{353\ {\rm GHz}}/S_{1.4\ {\rm GHz}} = 1.1 \times (1+z)^{3.8}
\end{equation}
The primary uncertainty in this relation is the dust temperature
dependence, which in a local ULIG sample produces a multiplicative factor 
of 2 range in the ratio relative to Arp~220. 
We have found that all known submm sources with radio fluxes 
or limits and spectroscopic identifications are broadly consistent with 
the expected ratio, although those with AGN characteristics have slightly 
lower ratios.

\section{Summary}

About 30 per cent of the 850\ $\mu$m background is already resolved,
and the slope of the counts is sufficiently steep (a power law index
of -2.2 for the cumulative counts) that only a small extrapolation to
fainter fluxes will give convergence with the background. Thus,
the typical submm source contributing to the background seems to be
in the $1-2$\ mJy range. 

Identifying the optical/NIR counterparts to submm sources is difficult 
due to the poor submm spatial resolution and the intrinsic faintness 
of the sources. In a spectroscopic survey of a lensed
submm sample, only about one-quarter of the sources had optical
counterparts bright enough to be spectroscopically identified,
a large fraction of which showed AGN characteristics.
We have recently found an alternative method for locating submm 
sources by targeting with SCUBA optical/NIR-faint radio sources whose 
positions are known to subarcsecond accuracy.
Our success with this method suggests that a large fraction of 
the sources in submm surveys have 
extremely faint optical/NIR ($K=21-22$) counterparts
and hence cannot be followed up with optical spectroscopy. However,
redshift estimations can be made for these sources using the 
submm-to-radio ratios. We find that the detected sources fall in the 
same $z=1-3$ range as the spectroscopically identified sources.
While still preliminary, the results suggest that the submm
population dominates the star formation in this redshift
range by almost an order of magnitude over the mostly distinct
populations selected in the optical/ultraviolet.

% As per the table environment, within the figure environment the \caption
% command should contain only the caption text.  The "Figure N." identification
% is generated by the \caption command on its own.
%
% Some appropriate amount of vertical space is opened up with the \vspace
% command.  Your figure has to fit in this space when you glue it in.
%
% You cannot use footnotes within figures.

% Finally, we have a little acknowledgements section.

\acknowledgments
We thank our collaborators Dave Sanders, 
Ian Smail, Rob Ivison, Andrew Blain, and Jean-Paul Kneib for
contributions to the work presented here.

% That's the end of the main body of the paper.  Now we will have some
% back matter.

% Now comes the reference list.  Since we typed out the citations ourselves,
% the reference list is enclosed in a "references" environment.  Each
% new reference begins with a \reference command which sets up the proper
% indentation.  Typography that may be required in the reference list by
% the editorial staff must be included by the author.
%
% Observe the "standard" order for bibliographic material: author name(s),
% publication year, journal name, volume, and page number for articles.
% Some journal names are available as macros; see the WGAS markup
% instructions for a listing of which ones have been "macro-ized".
% Note the use of curly braces to delimit the font changes: it is essential
% that this be done to limit the scope of the font declaration.
%
% There is no need to engage in any other typographic manipulation.

\end{document}